\documentclass{iauc}
\usepackage{graphicx}
\pubyear{2005}
\volume{199}
\pagerange{1--}
\jname{Probing Galaxies through Quasar Absorption Lines}
\editors{P. R. Williams, C. Shu, and B. M\'{e}nard, eds.}
\title[Ly$\alpha$ absorbers and Broadband Luminosities] % give here short title%
{The Stochastic Effect of Ly$\alpha$ absorbers on Broadband Luminosities of
Galaxies}

\author[T. Tepper-G., U. Fritze-v.A.]   %% give here short author list %%
{Thorsten Tepper-Garc\'\i a$^1$%
  \thanks{Present address: Uni-Sternwarte G\"ottingen, Geismarlandstr.
 11, 37083 G\"ottingen, Germany.},
\break \and Uta Fritze-v. Alvensleben$^1$}

\affiliation{$^1$ Uni-Sternwarte G\"ottingen, Geismarlandstr. 11, 37083
G\"ottingen, Germany.\break
email: tepper, ufritze@astro.physik.uni-goettingen.de}

\begin{document}

\maketitle

\begin{abstract}
We investigate the variations of broadband luminosities of low and
in\-ter\-me\-diate redshift galaxies due to the stochastic nature of the neutral
hydrogen distribution present in form of Ly$\alpha$ absorbers in the
intergalactic medium. This effect is caused by variations in the distribution
and properties (redshift, column density, Doppler parameter) of the absorbers
along different lines-of-sight out to a given redshift. Using a set of
observationally constrained redshift-, column density- and Doppler parameter
distributions we perform Monte Carlo simulations for a large number of
lines-of-sight towards galaxies at a given redshift $z_{em}$ and calculate
attenuated FUV/NUV magnitudes and corresponding 1-, 2- and 3$\sigma$ variations.
\textbf{We predict significant variations in luminosity ranging from -0.22 to
+0.48 magnitudes at the 1$\sigma$ level for galaxies at $z_{em}=1.5$ in
\textit{GALEX} data.}

%% add here a maximum of 10 keywords, to be taken form the file <Keywords.txt>.

\end{abstract}

\firstsection % if your document starts with a section,
              % remove some space above using this command.

\section{Introduction}

The observed photometric properties of local and distant galaxies are important
to constrain evolutionary synthesis models describing galaxy formation and
evolution. But estimating colors and luminosities of galaxies from models and
comparing them to the observed ones is a difficult task since, among other
effects, the light of a galaxy is partially absorbed by the intervening neutral
hydrogen (H{\sc i}) present in the intergalactic medium (IGM) in form of discrete
\emph{Ly$\alpha$ absorbers}. The distribution of these Ly$\alpha$ absorbers
along the line-of-sight (LOS) is stochastic in nature, \textit{i.e.} the
population of Ly$\alpha$ absorbers characterized by their redshift $z_{abs}$,
column density $N_{\mathrm{H{\sc i}}}$ and Doppler parameter $b$ is unique for
each LOS. Since the observation of individual galaxies implies necessarily the
observation along different lines-of-sight, we expect that the stochastic nature
of the Ly$\alpha$ absorbers distribution, especially of those at the highest
column densities, should cause a significant scatter in the observed broadband
colors, even for galaxies with similar intrinsic colors. Furthermore, we expect
this stochastic effects to be strongest along the shortest LOSs and, hence, to
be only observable in the UV as \textit{e.g.} with \textit{GALEX}.

In order to correctly account for the absorption of light caused by this
intervening absorbing H{\sc i}-systems, a detailed knowledge of their properties
(\textit{i.e.} $z_{abs}, \ N_{\mathrm{H{\sc i}}} , \ b$) is required. It has
been found from statistical analyses of quasar absorption lines that, when
averaged over a large number of LOSs, the mean distribution and properties of
Ly$\alpha$ absorbers are well described by a redshift- and column
density-dependent power law of the form of eq. \ref{eq:ddf} (see \textit{e.g.}
\cite{kim97} and references therein). Using different sets of parameters for
this power law derived from observations, there have been various studies (see
\textit{e.g.} \cite{moe90}, \cite{mad95}, \cite{BCG99}, here after
\cite*{BCG99}), which, with a different purposes and approaches, have made an
effort to account for the mean attenuation of the light of distant galaxies due
to intervening Ly$\alpha$ absorbing systems and the dispersion around this mean.
In this work, we make a further effort to calculate the stochastic effect of
the Ly$\alpha$ absorbers on the photometric properties of galaxies. Applying
some of the methods developed by \cite{moe90} and \cite*{BCG99}, we recover the
formalism of \cite{mad95}, \textit{i.e.} we model the mean attenuation and any
desired $\sigma$ level around this mean as a function of redshift. The advantage
of our approach  is that it is of general use in that it is completely
\emph{independent} of the galaxy input spectrum, and that it includes the
corrections concerning the calculation of the $\sigma$ scatter pointed out by
\cite*{BCG99}. As an application of our model, we estimate mean magnitudes and
scatter at the 1-, 2- and 3$\sigma$ level in the \textit{GALEX} FUV/NUV
broadband filters for galaxies with flat input spectra at $0.3 < z_{em} < 1.5$.

\section{The Model}

We present Monte Carlo simulations for the distribution of intervening
H{\sc i}-systems along random LOSs out to a given redshift $z_{em}$ (see also
\cite{moe90}, \cite*{BCG99}). As input we use differential distributions for the
number density of clouds as a function of redshift and their column densities of
the form
\begin{equation} \label{eq:ddf}
\frac{\partial^2 \mathcal{N}}{\partial N_{\mathrm{H{\sc i}}} \ \partial z}
= \mathcal{N}_0 \cdot \left ( 1 + z \right )^{\gamma} \cdot
N_{\mathrm{H{\sc i}}}^{-\beta}
\end{equation}
and a redshift-dependent Gaussian distribution for the Doppler parameters, all
derived from observations (\cite{kim97}, \cite*{BCG99}). We generate an ensemble
of $4 \cdot 10^{3}$ LOSs out to a given $z_{em}$, each of them containing a
random absorber population drawn from the input distributions. For each absorber
along each LOS and we calculate the photoelectric and Lyman-Series absorption
assuming Voigt-profiles and a \textit{flat} input spectrum with rest-frame
resolution \mbox{$\Delta \lambda = 0.5$ \AA}.
%(see \ref{fig:los})
%
%\begin{figure}[!h]
%\begin{center}
%\includegraphics[scale=0.3, angle=-90]{los2}
%\caption{Example of an attenuated spectrum along a single random LOS out to
%$z_{em} = 2.0$ (top left); Histogram of the clouds' properties along this
%particular LOS: redshift (bottom left), Doppler parameter (top right), and
%column density (bottom right).} \label{fig:los}
%\end{center}
%\end{figure}
%
We obtain a set of transmission values at each point of our wavelength baseline
and get in this way an ensemble of transmission values in the range $[0,1]$ as a
function of wavelength.

On this basis we define the Transmission Function \mbox{$T(e^{-\tau} \mid
\lambda\,;z_{em})$} as the residual transmission along a random LOS out to
$z_{em}$ at a given wavelength $\lambda$ as a function of the absorption
coefficient $e^{-\tau}$. It gives, up to a normalization constant, the
probability for the absorption coefficient $e^{-\tau}$ to take a certain value
in the range $[0,1]$ at a given wavelength along a random LOS out to $z_{em}$.
Since there is no known analytic expression for \mbox{$T(e^{-\tau} \mid
\lambda\,;z_{em})$}, we determine it numerically from our simulations. It turns
out that this distribution function is rather peculiar, and the calculation of
relevant statistical moments for the absorption assuming Gaussian statistics may
lead to incorrect estimates, as already noted by \cite*{BCG99}.

\begin{figure}[!h]
\begin{center}
\includegraphics[scale=0.3, angle=-90]{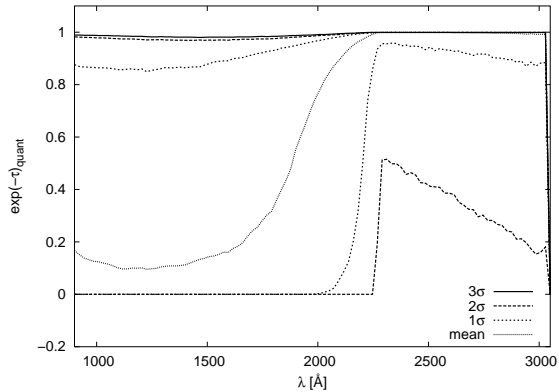}
\caption{Attenuation curves for $z_{em}=1.5$: mean and 1-, 2- and 3$\sigma$
variations (see eq. \ref{eq:qua}). For clarity, the curve corresponding to 'mean
- 3$\sigma$' is not shown.} \label{fig:att}
\end{center}
\end{figure}

Let us define $\tau_{\textnormal{q}}$ as the q transmission quantile via the
integral equation
\begin{equation} \label{eq:qua}
q = C_0 \int_0^{\tau_{\textnormal{q}}} T(1 - \tau \mid
\lambda\,;z_{em}) \, \mathrm{d}\tau
\end{equation}
where $C_0$ is a normalization constant, and the mean value of
\mbox{$T(e^{-\tau} \mid \lambda\,;z_{em})$} as $\tau_{50}$; and the 1-, 2- and
3$\sigma$ ranges around this value as $\tau_{50\pm34.13}$, $\tau_{50\pm47.72}$,
and $\tau_{50\pm49.86}$, respectively. Using this definition we calculate mean
attenuation curves (see \cite{mad95}) and the corresponding 1-, 2-, and
3$\sigma$ levels for any desired redshift, which can be applied to a given 
spectrum to estimate the amount of absorption due to the stochastic distribution
of the Ly$\alpha$ absorbers and their properties (see fig.\ref{fig:att}).

\section{Preliminary Results \& Conclusions}

\begin{table}
\begin{center}
\begin{tabular}{c c c c c c}
\hline
$z_{em}$ & Passband & mean &  $1\sigma$ &  $2\sigma$  & $3\sigma$\\
\hline
0.3 & FUV & -18.82 & ${}^{-}_{-}$ & ${}^{+0.13}_{\phantom{-}-}$ &
${}^{+1.16}_{\phantom{-}-}$\\[3pt]
0.3 & NUV & -20.26 & ${}^{-}_{-}$ & ${}^{+0.02}_{\phantom{-}-}$ &
${}^{+0.13}_{\phantom{-}-}$\\[3pt]
0.5 & FUV & -18.69 & ${}^{+0.02}_{\phantom{-}-}$ & ${}^{+0.28}_{\phantom{-}-}$ &
${}^{-}_{-}$\\[3pt]
0.5 & NUV & -20.25 & ${}^{-}_{-}$ & ${}^{+0.04}_{\phantom{-}-}$ &
${}^{+0.20}_{\phantom{-}-}$\\[3pt]
1.0 & NUV & -20.25 & ${}^{+0.19}_{-0.01}$ & ${}^{+0.52}_{-0.01}$ &
${}^{+1.27}_{-0.01}$\\[3pt]
1.5 & NUV & -20.02 & ${}^{+0.48}_{-0.22}$ & ${}^{+1.56}_{-0.24}$ &
${}^{\phantom{-}-}_{-0.24}$\\	
\hline
\end{tabular}
\caption{Mean and 1-, 2- and 3$\sigma$ variations in luminosity
($\mathrm{M_{AB}}$) in the \textit{GALEX} FUV and NUV passbands for a flat input
spectrum at four different redshifts \mbox{$z_{em} \in \lbrace 0.3, 0.5, 1.0,
1.5 \rbrace$}. In the cases where no value is given, either there is no
significant variation or the flux inside the given filter has dropped to zero
and the corresponding luminosity is not longer defined, being the case when the
Lyman-break has already cross the filter or due to strong line absorption in the
given filter.}
\end{center}
\end{table}
We apply our method to calculate the effect of the Ly$\alpha$ absorbers on the
FUV/NUV-broadband luminosities of a galaxy at redshifts $z \in \lbrace 0.3, 0.5,
1.0, 1.5 \rbrace$. Our results are given in Table 2. A careful inspection of
these results shows that the contingent presence of Ly$\alpha$ clouds with the
highest column densities can cause \textbf{substantial variations in
luminosity} already at the 1$\sigma$ level, ranging from -0.22 to +0.48
magnitudes.

\begin{acknowledgments}

T. T. G. thanks the IAU, the \textit{Astronomische Gesellschaft} (AG), and the
\textit{Berliner-Ungewitter Stiftung} (BU) for partial travel support.

\end{acknowledgments}

\end{document}